%Paper: hep-ph/9402231
%From: rmohapatra@umdhep.umd.edu
%Date: Fri, 04 Feb 1994 15:51:41 EST
%Date (revised): Wed, 09 Feb 1994 11:18:38 EST

% edited by xxx admin so that it would TeX automatically

\input jnl

\def\apj{\journal Ap.\ J., }

\def\nature{\journal Nature, }

\def\np{\journal Nucl.\ Phys., }

\def\pl{\journal Phys.\ Lett., }

\def\prd{\journal Phys.\ Rev.\ D, }

\def\prl{\journal Phys.\ Rev.\ Lett., }

\def\rmp{\journal Rev.\ Mod.\ Phys., }

\def\zphys{\journal Z.\ Phys., }
\def\gtwid{\mathrel{\raise.3ex\hbox{$>$\kern-.75em\lower1ex\hbox{$\sim$}}}}
\def\ltwid{\mathrel{\raise.3ex\hbox{$<$\kern-.75em\lower1ex\hbox{$\sim$}}}}

\rightline{UCSB--HEP--94-03}
\rightline{UMD-PP-94-90}
\rightline{February,1994}
\vskip 5mm

\centerline{\bf Accommodating solar and atmospheric neutrino deficits,}
\centerline{\bf hot dark matter, and a double beta decay signal}
\vskip 1cm
\centerline{{\bf David O.\ Caldwell}\footnote*{Supported in part
 by the Department of Energy}}
\centerline{Department of Physics, University of California, Santa Barbara,
California 93106}
\vskip 5mm
\centerline{{\bf Rabindra N.\ Mohapatra}+{Supported
 in part by the National Science Foundation}}
\centerline{Department of Physics, University of Maryland, College Park,
Maryland 20742}
\vskip 1cm
\centerline{ ABSTRACT}

{\singlespace\narrower\narrower
Neutrino mass explanations of the solar and atmospheric neutrino deficits and a
hot dark matter component require one of three patterns of those masses, as
already pointed out by us.  Recently there have been indications of a
non-vanishing amplitude for neutrinoless double beta decay.  If this additional
hint of neutrino mass is true, it would make even less likely the one unfavored
pattern (a sterile neutrino giving warm, rather than hot dark matter), would
alter another by making the $\nu_e$ a contributor to the hot dark matter, and
would make the third ($\nu_e$, $\nu_\mu$, and $\nu_\tau$ approximately
degenerate) much more likely than previously.  For this third case we construct
a gauge model consistent with other weak interaction data.  This model utilizes
a more general version of the see-saw mechanism, which is very likely to be the
source of neutrino mass, if this degenerate pattern is correct.
A new supernova constraint is utilized, and implications and tests
of the different mass matrices are noted.}

\vfill\eject
\oneandahalfspace
\centerline{I.\ INTRODUCTION}
There are several different observations involving neutrinos which receive a
plausible and satisfactory explanation if the neutrinos are massive, which they
are not in the Standard Model.  First is the well-known solar neutrino deficit
[1], observed by four different experiments [2].  Second is the deficit of muon
neutrinos relative to electron neutrinos produced in the atmosphere, as
measured by three experiments [3].  Third is the likely need for a neutrino
component of the dark matter of the universe to understand the structure and
density on all distance scales [4].  We showed [5] that, consistent with
particle physics and cosmological constraints, there are only three
conceivable patterns of neutrino mass which could explain these three
phenomena.  Of the three, one gave warm dark matter, rather than the favored
hot component, and one other also appeared dubious for Majorana masses because
of the limitation on electron neutrino mass from neutrinoless double beta
decay, $\beta\beta_{0\nu}$.

The $\beta\beta_{0\nu}$ situation has now changed, however, and leads us to
emphasize this formerly less favored pattern of neutrino masses.  What was
previously [6] a limit on effective neutrino mass has, after another year and a
half of
data taking, become about a two-standard-deviation indication for that mass
[7].  With current matrix element calculations, this effective Majorana mass
from the enriched $^{76}$Ge experiment is $\langle m_\nu\rangle\sim1$--2 eV.
Adding interest to this possibility is the observation in a $^{130}$Te
experiment [8] of a similar two-standard-deviation effect.  In this case the
even more uncertain matrix element calculations would favor
$\langle m_\nu\rangle\sim4$ eV.  Clearly much more experimental work is needed
before this hint of neutrino mass is even at the level of believability of the
three mentioned above, but the uncertainties in the nuclear matrix elements
make it possible that the $^{76}$Ge and $^{130}$Te experiments are observing
the same real effect.  In case this is true, we wish to point out how all of
these results could be accommodated theoretically.

First, however, we review the constraints on neutrino mass from experiments,
cosmology, and astrophysics.

\centerline{II. INPUT INFORMATION}
\centerline{A. Solar Neutrino Deficit}
For massive neutrinos which can oscillate from one species to another, the
solar electron neutrino observations [2] can be understood if the neutrino mass
differences and mixing angles fall into one of the following ranges [9], where
the Mikheyev-Smirnov-Wolfenstein (MSW) mechanism is included [10]:
$$\eqalignno{
  {\rm a)}&{\rm Small-angle\ MSW,\ }\Delta m^2_{ei}\sim6\times10^{-6}{\rm
eV}^2,
         \ \sin^22\theta_{ei}\sim7\times10^{-3},&\cr
  {\rm b)}&{\rm Large-angle\ MSW,\ }\Delta m^2_{ei}\sim9\times10^{-6}{\rm
eV}^2,
         \ \sin^22\theta_{ei}\sim0.6,&(1)\cr
  {\rm c)}&{\rm Vacuum\ oscillation,\ }\Delta m^2_{ei}\sim10^{-10}{\rm eV}^2,
         \ \sin^22\theta_{ei}\sim0.9.&\cr}$$

Of these, (a) is favored over (b) by the fits to the solar neutrino data [9],
and both (b) and (c) are disfavored by information from the neutrino burst from
supernova 1987A [11].

\centerline{B. Atmospheric Neutrino Deficit}
The second set of experiments indicating non-zero neutrino masses and mixings
has to do with atmospheric $\nu_\mu$'s and $\nu_e$'s arising from the decays of
$\pi$'s and $K$'s and the subsequent decays of secondary muons produced in the
final states of the $\pi$ and $K$ decays.  In the underground experiments the
$\nu_\mu$ and ${\bar\nu}_\mu$ produce muons and the $\nu_e$ and ${\bar\nu}_e$
lead to $e^\pm$.  Observations of $\mu^\pm$ and $e^\pm$ indicate a far lower
value for $\nu_\mu$ and ${\bar\nu}_\mu$ than suggested by na\"\i{}ve counting
arguments which imply that $N(\nu_\mu+{\bar\nu}_\mu)=2N(\nu_e+{\bar\nu}_e)$.
More precisely, the ratio of $\mu$ events to $e$-events can be normalized to
the ratio of calculated fluxes to reduce flux uncertainties, giving [3]
$$\eqalign{R(\mu/e)&=0.60\pm0.07\pm0.05\ {\rm (Kamiokande)},\cr
                   &=0.54\pm0.05\pm0.12\ {\rm (IMB)},\cr
                   &=0.69\pm0.19\pm0.09\ {\rm (Soudan\ II)}.\cr}$$

Combining these results with observations of upward going muons by Kamiokande
[3], IMB [3], and Baksan [12] and the negative Fr\'ejus [13] and NUSEX [14]
results leads to the conclusion [15] that neutrino oscillations can give an
explanation of these results, provided
$$\Delta m^2_{\mu i}\approx0.005{\ \rm to\ 0.5\ eV}^2,\ \sin^22\theta_{\mu i}
                    \approx0.5.\eqno(2)$$

\centerline{C. Hot Dark Matter}
There is increasing evidence that more than 90\% of the mass in the universe
must be detectable so far only by its gravitational effects.  This dark matter
is likely to be a mix of $\sim30$\% of particles which were relativistic at the
time of freeze-out from equilibrium in the early universe (hot dark matter) and
$\sim70$\% of particles which were non-relativistic (cold dark matter).  Such a
mixture [16] gives the best fit [4] of any available model to the structure and
density of the universe on all distance scales, such as the anisotropy of the
microwave background, galaxy-galaxy angular correlations, velocity fields on
large and small scales, correlations of galaxy clusters, etc.  A very plausible
candidate for hot dark matter is one or more species of neutrinos with total
mass of $m_{\nu_H}=93h^2F_H\Omega=7$ eV, if $h=0.5$ (the Hubble constant
in units of 100 km$\cdot$s$^{-1}\cdot$Mpc$^{-1}$), $F_H=0.3$ (the fraction
of dark matter which is hot), and $\Omega=1$ (the ratio of density of the
universe to closure density).  We shall use the frequently quoted 7 eV below,
but different determinations give $h=0.45\pm0.09$ [17] or $h=0.80\pm0.11$ [18]
(a value giving difficulties with $\Omega=1$), making
$m_{\nu_H}=2$ or 21 eV.  However, $F_H$ is probably less than 0.3, since a
baryonic content of $F_B=(0.010-0.015)/h^2$ must be accommodated and $F_H=0.2$
is not unlikely [19].

It is usually assumed that the $\nu_\tau$ would supply the hot dark matter.
This is justified on the basis of an appropriately chosen see-saw model [20]
and a $\nu_e\to\nu_\mu$ MSW explanation of the solar $\nu$ deficit.  However,
if the atmospheric $\nu_\mu$ deficit is due to $\nu_\mu\to\nu_\tau$, the
$\nu_\tau$ alone cannot be the hot dark matter, since the $\nu_\mu$ and
$\nu_\tau$ need to have close to the same mass.  It is interesting that instead
of a single $\sim7$ eV neutrino, sharing that $\sim7$ eV between two or among
three neutrino species provides a better fit to the universe structure and
particularly a better understanding of the variation of matter density with
distance scale [21].

\centerline{D. Nucleosynthesis Limits on Neutrino Species}
While the $Z^0$ width limits the number of weakly interacting neutrino species
to three, the nucleosynthesis limit [22] of about 3.3 on the number of light
neutrinos is more useful here, since it is independent of the neutrino
interactions.  Invoking a fourth neutrino, $\nu_s$, which is sterile, meaning
it does not have the usual weak interaction, must be done with parameters such
that it will not lead to overproduction of light elements in the early
universe.  For example, the atmospheric $\nu_\mu$ problem cannot be explained
by $\nu_\mu\to\nu_s$, since $\sin^22\theta_{\mu s}\approx0.5$ is too large for
the $\Delta m^2_{\mu s}$ involved, and that $\nu_s$ would have been brought
into equilibrium in the early universe [23].  On the other hand, the solar
$\nu_e$ problem can be explained by $\nu_e\to\nu_s$ for either the small-angle
MSW or the vacuum oscillation solutions, but not for the less favored
large-angle MSW solution [23].

\centerline{E. Neutrinoless Double Beta Decay}
As mentioned in the Introduction, there are indications from the
Heidelberg-Moscow $^{76}$Ge experiment [7] and the Milan $^{130}$Te experiment
[8] that there may be an effective Majorana neutrino mass of
$$\langle m_\nu\rangle\sim1-2\ {\rm eV}.\eqno(3)$$
In terms of individual neutrino masses $m_{\nu_i}$ and mixing matrix elements
$U_{ij}$,
$$\langle m_\nu\rangle\approx|\sum_i\eta_iU^2_{ei}m_{\nu_i}|,\eqno(4)$$
where $\eta_i=\pm1$, depending on the CP property of the individual neutrino.
Thus there may be a cancellation in the sum, making it possible that the
$\langle m_\nu\rangle$ which $\beta\beta_{0\nu}$ can measure is much smaller
than the actual $m_{\nu_e}$.  We shall address this issue below.

\centerline{III. POSSIBLE PATTERNS OF NEUTRINO MASS}
\centerline{A. Patterns Required by Solar and Atmospheric Neutrino Deficits and
Hot Dark Matter}
With the above input information, if the solar neutrino puzzle is resolved by
either $\nu_e\to\nu_\mu$ or $\nu_e\to\nu_s$ oscillations, the atmospheric
neutrino deficit is due to $\nu_\mu\to\nu_\tau$ oscillations, and some hot dark
matter is required, then there are only three possible textures for neutrino
masses, as we have pointed out [5]:
\item{(i).} All three neutrinos are nearly degenerate, with $m_{\nu_e}\approx
m_{\nu_\mu}\approx m_{\nu_\tau}\approx2$--3 eV, since $\nu_e\to\nu_\mu$
and $\nu_\mu\to\nu_\tau$ both require small mass differences, but the required
dark matter mass can be shared.
\item{(ii).} The three neutrinos with weak interactions are light, and a
sterile neutrino supplies the dark matter, but the early decoupling required to
satisfy the nucleosynthesis constraint reduces the number density of the
$\nu_s$, forcing the mass of $\nu_s$ to be so large that it becomes warm,
rather than the desired hot, dark matter.
\item{(iii).} The $\nu_e$ and $\nu_s$ can be quite light to take care of the
solar neutrino problem while the $\nu_\mu$ and $\nu_\tau$ share the dark matter
role, being $\sim3$--4 eV each, and explain the atmospheric $\nu_\mu$ deficit.

\centerline{B. Added Effect of a Non-Zero Neutrinoless Double Beta Decay
Amplitude}
The primary purpose of this paper is to discuss the impact on the neutrino mass
textures worked out in Ref.\ [5] for the above three cases should the
indications for an effective Majorana mass for the neutrino, $\langle
m_\nu\rangle\sim1$--2 eV, be confirmed [24].  First, however, it is worth
pointing out that there could then be no quadratic or linear see-saw model
justification for the $\nu_\tau$ alone to be hot dark matter.

If convincing peaks are seen at the appropriate energies for more than one
$\beta\beta$ isotope, and $\langle m_\nu\rangle$ is determined within the
uncertainties of the nuclear matrix elements, there still remains the issue of
possible cancellations in Eq.~(4) in trying to extract the mass of $\nu_e$ (or
more precisely, $\nu_1$).  As was pointed out in [5], for case (i) the
unitarity of the mixing matrix limits the one undetermined mixing angle, that
of $\nu_e$--$\nu_\tau$, to 0.05 for the small-angle MSW solution.  The
smallness of the $\nu_e$--$\nu_\tau$ and $\nu_e$--$\nu_\mu$ angles and hence of
the mixing matrix elements $U_{e2}$ and $U_{e3}$ makes $\langle
m_\nu\rangle\approx m_{\nu_1}\approx m_{\nu_e}$ for this case.  This conclusion
is true also for the large-angle MSW and vacuum oscillation solutions, since
the near mass degeneracy of the $\nu_1$, $\nu_2$, and $\nu_3$ forces the
$\nu_2$ and $\nu_3$ terms in Eq.~(4) to nearly cancel, or else
$\beta\beta_{0\nu}$ would have been observed long ago.

For case (iii), as in case (i), the same considerations would apply for the
small-angle MSW solution, since the $\nu_s$ has to be very weakly mixed to
satisfy the nucleosynthesis bound.  Nucleosynthesis also eliminates the
large-angle MSW solution and in the vacuum oscillation case forces the $\nu_s$
to be mixed strongly only with the $\nu_e$.  In this last instance (case (iii),
vacuum oscillations), a limitation on the effect of the $\nu_\mu$ and
$\nu_\tau$ masses may be invoked from supernovae considerations.  If, as
appears likely, the heavy elements in the universe are produced by a rapid
neutron capture process in the supernova environment, then the
$\nu_e$--$\nu_\mu$ and $\nu_e$--$\nu_\tau$ mixing angles are severely
restricted ($\sin^22\theta\ltwid4\times10^{-4}$) for $\Delta m^2\gtwid4$ eV$^2$
(with a rapid decrease in $\sin^22\theta$ for larger $\Delta m^2$) [25].
Otherwise the energetic $\nu_\mu$ and $\nu_\tau$ ($\langle E\rangle\approx25$
MeV) can convert to $\nu_e$'s which have much higher energy than the thermal
$\nu_e$'s ($\langle E\rangle\approx11$ MeV).  The higher energy $\nu_e$'s,
having a larger cross section, will reduce the neutron density via $\nu_e+n\to
e^-+p$, diminishing heavy element formation.  With the effect of the $\nu_\mu$
and $\nu_\tau$ eliminated in Eq.~(4) and $\nu_e$ and $\nu_s$ of approximately
equal mass, the pseudo-Dirac (opposite CP eigenvalues) combination is not
allowed, since $\langle m_\nu\rangle\approx0$, whereas in the case of the same
CP eigenvalues, $m_{\nu_e}\approx\langle m_\nu\rangle/2$.

The added $\beta\beta_{0\nu}$ constraint makes case (ii) even less likely but
now for a different reason.  Since the $\nu_e$, $\nu_\mu$, and $\nu_\tau$ could
provide much, or even all, of the hot dark matter, there is little or no
reason to invoke a sterile neutrino.  We will not deal any further with this
unlikely possibility.

In case (iii), the new information alters the allowed range of parameters.  For
example, $\nu_e$ and $\nu_s$, recalling the uncertainty in the
$\beta\beta_{0\nu}$ nuclear matrix elements, could be $\sim1$ eV each, but only
the $\nu_e$ would contribute to dark matter.  The $\nu_s$ must decouple early
(probably even before the quark-hadron phase transition at $T\sim200$ MeV), in
order not to contribute excessively (i.e., $\delta N_\nu\le0.3$ [22]) to the
energy density of the universe at the epoch of nucleosynthesis.  Using the
nominal 7 eV, the $\nu_\mu$ and $\nu_\tau$ would then be $\sim3$ eV each, and
the most likely way to determine that this pattern of masses is correct is to
observe $\nu_\mu\to\nu_e$ oscillations for $\Delta m^2\approx8$ eV$^2$, such as
could be done in the current LSND experiment at Los Alamos, perhaps requiring
some uncertainty in the supernova constraint [25] just discussed.  While
$\beta\beta_{0\nu}$ is the only currently feasible way to determine the masses
in scheme (i), and $\nu_\mu\to\nu_e$ oscillations plus $\beta\beta_{0\nu}$ are
the way to get masses in scheme (iii), the choice between these two
alternatives could be made on the basis of whether the solar $\nu_e$ deficit is
due to $\nu_e\to\nu_\mu$ or $\nu_e\to\nu_s$.  The measurement of the charged to
neutral current ratio by the Sudbury Neutrino Observatory distinguishes
$\nu_e\to\nu_\mu$ from $\nu_e\to\nu_s$, but proving the latter (as opposed
to an astrophysical cause of the $\nu_e$ deficit) requires a measurement of
spectral distortions [9,26]\footnote*{An alternative which we do not
consider is the $\nu_e-\nu_s$ oscillation as the solution to the
solar neutrino deficit but $\nu_{\mu}-\nu_e$ oscillation for the atmospheric
case by making $\nu_e$ slightly heavier than $\nu_{\mu}$; the $\nu_{\tau}$
would then be the main contributor to hot dark matter. This is unlikely
because the observed absolute $\nu_e$ flux as a function of energy
agrees with calculations. Note that if $m_{\nu_e}\simeq 2 eV$, this
becomes like case (iii) with a rather bizzare mass ordering.}.

\centerline{C. Mass Matrix for Case (iii)}
The generic form of the Majorana mass matrix for case (iii), as given in
Ref.~[5], in the basis ($\nu_s$, $\nu_e$, $\nu_\mu$,
$\nu_\tau$),
$$M=\pmatrix{\mu_1&\mu_3&\epsilon_{11}&\epsilon_{12}\cr
             \mu_3&\mu_2&\epsilon_{21}&\epsilon_{22}\cr
             \epsilon_{11}&\epsilon_{21}&m&\delta/2\cr
             \epsilon_{12}&\epsilon_{22}&\delta/2&m+\delta\cr}.\eqno(5)$$
For simplicity, we set the $\epsilon_{ij}=0$, and we shall demonstrate later
that those terms are indeed negligible for most purposes.  The implication of
the $\beta\beta_{0\nu}$ constraint is that $\mu_1+\delta_1=\mu_2=1$--2 eV.  For
the small-angle MSW case, $\mu_1\gg\delta_1\gg\mu_3\approx5\times10^{-8}$ eV,
and $\delta_1\approx 1.5\times 10^{-6}$ eV.  For the vacuum oscillation case,
$2\mu_3\approx\delta_1\approx10^{-10}$ eV.  Recall that the
large-angle MSW and the pseudo-Dirac
$\nu_e$--$\nu_s$ vacuum oscillation cases are ruled out.  A total neutrino mass
of the nominal 7 eV implies that $m\approx2$ eV, and
$\delta\approx10^{-1}$--$10^{-3}$ eV, which is a slight change from the values
of the parameters given in Ref.~[5].

Returning now to the $\epsilon_{ij}$, if the $\epsilon_{ij}\ll\mu_{1,2}$ (or
$\mu_3$) and $m$, then they perturb the eigenvalues only slightly, but they
lead to mixings between $\nu_\mu$ and $\nu_\tau$ with $\nu_s$.  These
$\nu_\mu$--$\nu_s$ and $\nu_\tau$--$\nu_s$ mixings are severely constrained by
nucleosynthesis [23], implying that $\epsilon_{ij}\ltwid10^{-5}$--$10^{-6}$ eV,
since $\Delta m^2\sim{\rm few}\ eV^2$.  Somewhat weaker constraints also result
from the supernova argument [25] given above, which leads to
$\epsilon_{2j}\ltwid10^{-3}$ eV, since $\theta_{ej}\approx\epsilon_{2j}/m$.  It
is, therefore, justifiable to neglect $\epsilon_{ij}$, except for determining
the mixings.

\centerline{D. Mass Matrices for Case (i)}
We turn now to the case of three highly degenerate neutrino eigenstates, which
will be the favored scenario, if future $\beta\beta_{0\nu}$ results yield a
mass large enough to account for a third of the hot dark matter mass.  Since
all the values of $\Delta m^2_{ij}$ in this case are small (i.e., $\ltwid0.5$
eV$^2$), the mixing angles are not constrained by heavy element production by
supernovae [25].  In a subsequent section, we will present a gauge model which
will provide this mass degeneracy in a ``natural" manner (i.e., without any
arbitrary adjustment of parameters), guaranteed by a horizontal symmetry.

Here and in Ref.~[5] our conclusions about favored neutrino mass textures have
been based on the assumption of two-flavor neutrino oscillations.  The solar
and atmospheric neutrino data can be treated together using a three-flavor
oscillation scenario, but only the recent work by Kim and Lee [27] could be
made compatible with the need for hot dark matter and $\langle
m_\nu\rangle\sim2$ eV, although they did not consider this possibility.  They
assumed that all three neutrinos mix maximally and found mass differences
compatible with the two-flavor oscillation values given in Eq.~(1) for vacuum
oscillations and in Eq.~(2) above.  Their assumption of maximal mixing requires
a different mass matrix than that given in Ref.~[5] for the
three-degenerate-neutrino case.  Both possibilities are given below in forms
compatible with solar and atmospheric neutrino data, a Majorana
$m_{\nu_e}\sim2$ eV, and total mass sufficient for hot dark matter.

\noindent\underbar{Case A.} The mass matrix in the $\nu_e,\ \nu_\mu,\ \nu_\tau$
basis is the same as in Ref.~[5], since it corresponds to two-flavor
oscillations:
$$M=\pmatrix{m+\delta_1s^2_1&-\delta_1c_1c_2s_1 & -\delta_1c_1s_1s_2\cr
-\delta_1c_1c_2s_1&m+\delta_1c^2_1c^2_2+\delta_2s^2_2&
             (\delta_1-\delta_2)c_2s_2\cr
     -\delta_1c_1s_1s_2&(\delta_1-\delta_2)c_2s_2&
             m+\delta_1s^2_2+\delta_2c^2_2\cr},\eqno(6)$$
where $c_i=\cos\theta_i$ and $s_i=\sin\theta_i$, $m=2$ eV;
$\delta_1\simeq 1.5\times 10^{-6}$ eV;
 $\delta_2\simeq.2$ to .002 eV; $s_1\simeq0.05$; and
$s_2\simeq0.4$ for the small-angle MSW solution.  Note that it is a somewhat
more accurate version of the Eq.~(1) in Ref.~[5].  For the large-angle MSW or
vacuum oscillation solutions, $\delta$, $s_1$, and $s_2$ will be different.

\noindent\underbar{Case B.} The mass matrix in the same basis but for
three-flavor oscillations and maximal mixings [27] is
$$M=\pmatrix{
     m+\delta_1+\delta_2&-x\delta_1-x^2\delta_2&-x^2\delta_1-x\delta_2\cr
     -x^2\delta_1-x\delta_2&m+\delta_1+\delta_2&x\delta_1+x^2\delta_2\cr
     -x\delta_1-x^2\delta_2&x^2\delta_1+x\delta_2&m+\delta_1+\delta_2\cr},
     \eqno(7)$$
where $m=2$ eV, $\delta_1={1\over4}\times10^{-10}$ eV,
$\delta_2={1\over4}(10^{-2}-10^{-3})$ eV, and $x=e^{2\pi i/3}$, now for the
case of vacuum oscillations only.

In comparing cases A and B, we see that in case A, except for the high degree
of degeneracy of the diagonal terms, the other entries are hierarchical but not
related to each other.  That is, elements $M_{12}\ll M_{13}\ll M_{23}\ll
M_{ii}$.  On the other hand, in case B, all non-diagonal elements are the same
to leading order and all diagonal entries are exactly the same; i.e.,
$M_{11}=M_{22}=M_{33}$ and $|M_{ij}|=\delta_2$ for $i\ne j$.  Case B,
therefore, will be much harder to obtain in a natural manner for a gauge model
than Case A.  For this reason, and because the vacuum oscillation solution is
much less likely, in the next section we present a gauge model which reproduces
a mass matrix of the type in Case A only.

\centerline{IV. DEGENERATE MAJORANA NEUTRINOS FROM GRANDUNIFICATION}
Before discussing details of the model, we note some important points about any
gauge model that could reproduce the mass matrix of case A, Eq.~(6).  First,
the high degree of degeneracy of the neutrino masses, and no evidence for such
degeneracy elsewhere in the fermion spectrum, implies that the dominant masses
for neutrinos must arise from a different source than that of quarks and
charged leptons.  This feature is present in see-saw type models, as we show
below.  Second, there must be a horizontal symmetry, $G_H$, in the neutrino
sector, and the minimal $G_H$ (whether discrete or continuous) must have a
three-dimensional representation.  Third, since the horizontal symmetry is
broken in the quark and charged lepton masses, one expects corrections to the
neutrino degeneracy from them, as well as any possible horizontal symmetry
breaking effects in the neutrino sector.  The smallness of the neutrino
degeneracy breaking terms must be stable under radiative corrections.

Below we present an SO(10) grandunified model which can lead to the desired
pattern of neutrino masses without arbitrary fine tuning of parameters, but
first we give the qualitative reason this works.  In the early days of the
discussion of the see-saw formula for neutrino masses, it was pointed out [28]
that implementing that mechanism in the simplest left-right or SO(10) models
resulted in a $\nu_L$-$\nu_R$ mass matrix of the modified see-saw form:
$$\bordermatrix{&\nu_L&\nu_R\cr
                \nu_L&fv_L&m_{\nu_D}\cr
                \nu_R&m^T_{\nu_D}&fv_R\cr},\eqno(7)$$
where $f$ and $m_{\nu_D}$ are $3\times3$ matrices and $v_L\approx\lambda
M^2_{W_L}/v_R$.  The light neutrino mass matrix that follows from
diagonalizing Eq.~(7) is
$$m_\nu\approx f\lambda M^2_{W_L}/v_R-m_{\nu_D}f^{-1}m^T_{\nu_D}/v_R+\ldots
\eqno(8)$$
While both terms vanish as $v_R\to\infty$, the first term always dominates
over the second one for neutrino masses.  This negates the usual quadratic
formula (i.e., the second term) for neutrino masses.  It means that usual
see-saw quadratic mass formula is not realized in generic left-right or SO(10)
models, unless additional assumptions are
made, as has been pointed out [29] (e.g., decoupling the parity and SU(2)$_R$
scales in left-right symmetric models, or in the singlet Majoron models, etc.).

Accepting the more complex see-saw formula given above, it is clear that if a
symmetry dictates that $f_{ab}=f\delta_{ab}$, then the neutrino masses are
degenerate to leading order.  For $v_R\approx10^{13.5}$ GeV,
$f\lambda\approx1$, we get $m_{\nu_e}=m_{\nu_\mu}=m_{\nu_\tau}\approx1.5$
eV, $m^2_{\nu_\mu}-m^2_{\nu_e}\approx3m^2_c/10fv_R\approx10^{-4}/f$ eV$^2$,
and $m^2_{\nu_\tau}-m^2_{\nu_\mu}\approx3m^2_t/10fv_R\approx(2/f)(m_t/150$
GeV)$^2$ eV$^2$.  These mass differences are of the right order of magnitude to
explain the solar neutrino (via the MSW mechanism) and the atmospheric neutrino
puzzles, while the masses roughly give the dark matter.

It is also interesting to note that the B-L breaking scale of
$v_R\sim10^{13}$ GeV emerges naturally from constraints of
$\sin^22\theta_W$ and $\alpha_s$ in non-supersymmetric SO(10) grandunified
theories [30], enhancing the reason for an SO(10) scenario.  To guarantee the
neutrino degeneracy (i.e., $f_{ab}=f\delta_{ab}$), an extra SU(2)$_H$ family
symmetry [31] is imposed on the model.  This family symmetry will be broken
softly by terms in the Lagrangian of dimension two, so that departures from the
degeneracy in the neutrino sector are naturally small.

Turning now to some details of the model, the known fermions of each family are
assigned as usual to a $\{16\}$-dimensional spinor
representation of SO(10), and we denote them by $\psi_a$ ($a=$ family index,
going over 1,2,3).  We assume
that the $\psi_a$ transform as a triplet under SU(2)$_H$, which is taken as a
softly broken global family symmetry.  As to the Higgs sector,
the SO(10) symmetry is assumed to break down at the GUT scale, $M_U$, to the
left-right symmetric group SU(2)$_L\times{\rm SU}(2)_R\times{\rm
SU}(4)_C\times P\equiv G_{224P}$ via the non-zero vev of a $\{54\}$-dimensional
scalar multiplet.  We wish to emphasize that the existence of parity as a good
local symmetry below the GUT scale is important for the more general see-saw
formula, (8), to work [29].  The $G_{224P}$ symmetry is broken down to the
Standard Model at a scale $v_R$ by a $\{126\}$-dimensional Higgs multiplet,
denoted by $\Delta_0$.  Since $\langle\Delta_0\rangle$ breaks B-L symmetry, it
gives the heavy Majorana mass to the right-handed neutrino.  The $\Delta_0$ is
assumed to be an SU(2)$_H$ singlet in order to guarantee the $f$ matrix in
Eq.~(7) to be an identity matrix (see later).

We assume that there are six complex
$\{10\}$-dimensional Higgs multiplets denoted by $H_{ab}$, (with
$H_{ab}=H_{ba}$, where $a,b=1,2,3$) which transform like a
$\{1\}+\{5\}$-dimensional representation under SU(2)$_H$ symmetry.  We denote
the singlet by TrH and the $\{5\}$ by $H^5$.  Two Standard Model doublets in
each H-multiplet acquire vev's at the electroweak scale and are denoted by
$\kappa^u_{ab}$ and $\kappa^d_{ab}$.  In order to get a realistic charged
fermion spectrum (and avoid disastrous relations such as $m_s=m_\mu$ at
$M_U$), we need five more $\{126\}$-dimensional multiplets denoted by
$\Delta_{ab}$ (with $\Delta_{ab}$ symmetric and traceless in the indices $a$
and $b$), which acquire induced vev's of order the electroweak scale.  We keep
the $(\rm mass)^2$ of $\Delta_{ab}$ to be positive and of order the GUT scale,
so that they do not break B-L symmetry and thus do not contribute to Majorana
masses.

The part of the Higgs potential that is responsible for the weak scale vev's,
as well as the B-L breaking vev is
$$\eqalign{V^0_1&=-\mu^2_{ab}H^{5\dagger}_{ab}H^5_{ab}-\tilde\mu^2_{ab}
           H_{ab}H_{ab}+\lambda_1(TrH^\dagger H)^2+\lambda_2TrH^\dagger H
           H^\dagger H+\lambda_3TrH^\dagger HH^\dagger TrH\cr
           &+\lambda_R(\Delta^\dagger_0\Delta_0-v^2_R)^2
           +M^2_{ab}\Delta^\dagger_{ab}\Delta_{ab}+\lambda_3(Tr\Delta^\dagger
           \Delta)^2+\lambda_4Tr(\Delta^\dagger\Delta\Delta^\dagger\Delta)\cr
           &+\beta_1\Delta^\dagger_0\Delta_0\Delta_{ab}H_{ab}+\beta_2
           \Delta^\dagger_0\Delta_0\Delta_0TrH+\gamma\Delta_0\Delta_0
           Tr(H H)+{\rm h.c.}\cr}$$
(In writing this potential, we have assumed a softly broken Peccei-Quinn (PQ)
symmetry, as in Ref.\ [32].)  The $\lambda_R$ term generates the $\Delta_0$-vev
which breaks the B-L symmetry and also breaks $G_{224P}$ down to the Standard
Model.  The terms $\beta_1$ and $\beta_2$ generate the induced weak-scale vev's
for the (2, 2, 15) components of both $\Delta_0$ and $\Delta_{ab}$; we denote
these vev's by $v^u_{ab}$ and $v^d_{ab}$, respectively.  This induced vev
mechanism [32] has the advantage that it avoids any need for extra fine tunings
of the $\Delta$ masses.  In fact, in this model only the masses of $H_{ab}$
need to be fine tuned to the electroweak scale.  Note that the $\gamma$ term in
the above equation generates the $v_L$ term in $\Delta_0$, which causes the
approximate neutrino degeneracy.

The SO(10)$\times$SU(2)$_H$-invariant Yukawa coupling is given by
$${\cal L}_Y=h\psi_a\psi_bH_{ab}+f_0\psi_a\psi_a\bar\Delta_0+f_1\psi_a\psi_b
\bar\Delta_{ab}+{\rm h.c.}$$
The softly broken PQ symmetry prevents the coupling of $H^*$ to fermions.

The resulting charged fermion mass matrix becomes
$$\eqalign{M_{u,ab}&=h\kappa^u_{ab}+f_0v^u_0\delta_{ab}+f_1v^u_{ab}\cr
           M_{d,ab}&=h\kappa^d_{ab}+f_0v^d_0\delta_{ab}+f_1v^d_{ab}\cr
           M_{\ell,ab}&=h\kappa^d_{ab}-3f_0v^d_0\delta_{ab}-3f_1v^d_{ab}\cr
           M_{\nu D}&=h\kappa^u_{ab}-3f_0v^u_0\delta_{ab}-3f_1v^u_{ab}.\cr}$$
The neutrino mass matrix has the desired form given in Eq.~(6).  Since there
are 23 free parameters in the mass matrices, it is not surprising that the
quark masses and mixings and charged lepton masses can be reproduced easily.
It is worth emphasizing that, even though the particle spectra of the model may
appear quite dense, the low energy spectrum consists only of the particles of
the Standard Model, along with massive neutrinos.

This model has a prediction for the lifetime of the proton, which can be
obtained from Ref.~[30].  Ignoring heavy-particle threshold corrections and QCD
coupling uncertainties, one gets $\tau_P\approx10^{33}$ years, which would be
reachable by Super Kamiokande.  The existence of so many heavy particles,
however, is likely to introduce an uncertainty by one or two orders of
magnitude.

If future experiments bear out a degenerate light neutrino spectrum, this
detailed and quite complicated SO(10) model may not be the appropriate
description of the physics.  Its essential ingredient, the correct general
see-saw formula, [8], will almost surely be required to fit those data,
however.

\centerline{V. CONCLUSIONS}
The solar and atmospheric neutrino data, along with a need for some hot dark
matter, if all are due to neutrino mass, limit the texture of those masses to
one of three possibilities, probably only two of which seem viable.  The third
alternative, under the conditions of Ref.~[5], had a sterile neutrino, $\nu_s$,
which gave warm dark matter.  If further neutrinoless double beta decay,
$\beta\beta_{0\nu}$, experiments confirm a Majorana mass for the $\nu_e$ around
2 eV or so, the $\nu_s$ could become superfluous.  A $\nu_e$ mass found by
$\beta\beta_{0\nu}$ would contribute to the hot dark matter in the viable
alternative scheme in which the solar problem is solved by $\nu_e\to\nu_s$, but
the main contribution would come from the $\nu_\mu$ and $\nu_\tau$, with
$\nu_\mu\to\nu_\tau$ solving the atmospheric $\nu_\mu$ problem.  The final
alternative has the hot dark matter mass shared almost equally among $\nu_e$,
$\nu_\mu$, and $\nu_\tau$, with the solar neutrino deficit now being
$\nu_e\to\nu_\mu$.  This last scheme becomes particularly attractive if $\nu_e$
is $\sim2$ eV, and so we have presented a gauge model that reproduces this
degenerate neutrino spectrum without affecting the vastly nondegenerate masses
for charged leptons.  Should this be the true pattern of neutrino mass, even if
this model is not the correct description, its basis in a version of the
see-saw mechanism which is more generally correct than that usually invoked is
very likely to be the source of those masses.

\vskip 5mm

\centerline{ REFERENCES}

\def\apj{\journal Astrophys.\ J., }
\item{[1].} J.N.~Bahcall and M.H.~Pinsonneault, \rmp 64, 885, 1992.
\item{[2].} R.~Davis et al., Proceedings of the 21st International Cosmic Ray
Conference, Vol.~12, edited by R.J.~Protheroe (Univ.\ of Adelaide Press,
Adelaide, 1990), p.~143; K.S.~Hirata et al., \prd 44, 2241, 1991; A.I.~Abrazov
et al., \prl 67, 3332, 1991 and V.N.~Gavrin in TAUP 93 Workshop, Gran Sasso,
Italy, 1993 (unpublished); P.~Anselmann et al., \pl B285, 376, 1992 and {\bf
B314}, 445 (1993).
\item{[3].} K.S.~Hirata et al., \pl B280, 146, 1992; R.~Becker-Szendy et al.,
\prd 46, 3720, 1992; P.J.~Litchfield in International Europhysics Conference on
High Energy Physics, Marseille, France, 1993 (unpublished).
\item{[4].} E.L.~Wright et al., \apj 396, L13, 1992; M.~Davis, F.J.~Summers,
and D.~Schagel, \nature 359, 393, 1992; A.N.~Taylor and M.~Rowan-Robinson,
\journal ibid., 359, 396, 1992; R.K.~Schaefer and Q.~Shafi, Report No.\
BA-92-28, 1992 (unpublished) and \nature 359, 199, 1992; J.A.~Holtzman and
J.R.~Primack, \apj 405, 428, 1993; A.~Klypin et al., \apj 416, 1, 1993.
\item{[5].} D.O.~Caldwell and R.N.~Mohapatra, \prd 48, 3259, 1993.
\item{[6].} A.~Balysh et al., \pl B283, 32, 1992.
\item{[7].} A.~Piepke in International Europhysics Conference on
High Energy Physics, Marseille, France, 1993 (unpublished).
\item{[8].} E.~Garcia in TAUP 93 Workshop, Gran Sasso,
Italy, 1993 (unpublished).
\item{[9].} Of the many analyses, we use the recent results of N.~Hata and
P.~Langacker, Report No.\ UPR-0592T, 1993 (to be published in Phys.\ Rev.) and
P.I.~Krastev and S.T.~Petcov, Report No.\ SISSA 177/93/EP, 1993 (unpublished).
\item{[10].} S.P.~Mikheyev and A.Yu.~Smirnov, \journal Yad.\ Fiz., 42, 1441,
1985; L.~Wolfenstein, \prd 17, 2369, 1978; {\bf 20}, 2634 (1979).
\item{[11].} A.Yu.~Smirnov, D.N.~Spergel, and J.N.~Bahcall, Institute for
Advanced Study Report No.\ IASSNS-AST 93/15, 1993 (to be published in
Phys.\ Rev.\ D).
\item{[12].} M.M.~Boliev et al.\ in Proceedings of the 3rd International
Workshop on Neutrino Telescopes, Venice, Italy, 1991, edited by M.~Baldo-Ceolin
(Istitute Nazionale di Fisica Nucleare, Padova, 1991), p.~235.
\item{[13].} Ch.~Berger et al., \pl B245, 305, 1990; {\bf 227}, 489 (1989).
\item{[14].} M.~Aglietta et al., \journal Europhys.\ Lett., 15, 559, 1991.
\item{[15].} W.~Frati et al., \prd 48, 1140, 1993.
\item{[16].} Q.~Shafi and F..~Stecker, \prl 53, 1292, 1984.
\item{[17].} A.~Sandage et al., \apj 401, L7, 1992.
\item{[18].} G.H.~Jacoby et al., \journal PASP, 104, 599, 1992.
\item{[19].} U.~Seljak and E.~Bertschinger, MIT Report MIT-CSR-93-33, 1993
(unpublished).
\item{[20].} M.~Gell-Mann, P.~Ramond, and R.~Slansky in Supergravity, edited by
D.~Freedman et al.\ (1979); T.~Yanagida, in KEK Lectures, edited by O.~Sawada
et al.\ (1979); R.~Mohapatra and G.~Senjanovi\'c, \prl 44, 912, 1980.
\item{[21].} M.~Davis and J.~Primack (private communications); Y.P.~Ying et
al.\ Univ.\ of Arizona Report 93-0590, 1993 (to be published in Astronomy and
Astrophysics).
\item{[22].} T.P.~Walker et al., \apj 376, 51, 1991.
\item{[23].} P.~Langacker, Univ.\ of Pennsylvania Report No.\ UPR 0401T, 1989
(unpublished); R.~Barbieri and A.~Dolgov, \np B349, 743, 1991; K.~Enqvist,
K.~Kainulainen, and J.~Maalampi, \pl B249, 531, 1990; M.J.~Thomson and
B.H.J.~McKellar, \pl B259, 113, 1991; V.~Barger et al., \prd 43, 1759, 1991;
P.~Langacker and J.~Liu, \prd 46, 4140, 1992; X.~Shi, D.~Schramm, and
B.~Fields, \prd 48, 2563, 1993; J.\ Cline, \prl 68, 3137, 1992.
\item{[24].} For a recent discussion of $\beta\beta_{0\nu}$ and just the solar
neutrino problem, see S.T.~Petcov and A.Yu.~Smirnov, Report No.\ SISSA
113/93/EP, 1993 (unpublished).
\item{[25].} Y.-Z.~Qian et al., \prl 71, 1965, 1993.
\item{[26].} S.M.~Bilenky and C.~Giunti, University of Torino Report No.\ DFTT
62/93, 1993 (unpublished).
\item{[27].} C.W.~Kim and J.A.~Lee, Johns Hopkins Univ.\ Report
No.\ JHU-TIPAC-930023, 1993 (unpublished).
\item{[28].} R.N.~Mohapatra and G.~Senjanovi\'c, \prd 23, 165, 1981.
\item{[29].} D.~Chang and R.N.~Mohapatra, \prd 32, 1248, 1985.
\item{[30].} R.N.~Mohapatra and G.~Senjanovi\'c, \zphys 17, 53, 1983;
N.G.~Deshpande, R.~Keith, and P.B.~Pal, \prd 46, 2261, 1992.
\item{[31].} After this work was done, we were informed that a degenerate
neutrino spectrum from SU(2)$_H$ symmetry in the context of low-scale
electroweak models has been discussed by K.S.~Babu and S.~Pakvasa (private
communication).
\item{[32].} K.S.~Babu and R.N.~Mohapatra, \prl 70, 2845, 1993.

\endit